\documentstyle[aps,prb,psfig,twocolumn,floats]{revtex}
\begin{document}
\draft

\title{Superflow in $d$-wave superconductors}

\author{J. Ferrer, M. A. Gonz\'alez-Alvarez}
\address{
Departamento de F\'{\i}sica, Facultad de Ciencias, Universidad de Oviedo, 
E-33007 Oviedo, Spain}
\author{J. S\'anchez-Ca\~nizares}
\address{
Departamento de F\'{\i}sica Te\'orica de la Materia Condensada, C-V, 
Universidad Aut\'onoma de Madrid, E-28049 Madrid, Spain} 

\address{
\begin{minipage}[t]{6.0in}
\begin{abstract}
Superflow in a phenomenological tight-binding model for the superconducting
state of some High-temperature superconductors is discussed thoroughly. 
The formalism used is explicitly gauge-invariant and currents are
computed exactly within BCS theory, going therefore beyond linear response
theory. The dependence of gap functions, current density, critical currents and 
free energy as a function of the superfluid velocity for different angles and 
doping concentrations is investigated. Different sources of anisotropy, like 
the dispersion relation of the model, the internal symmetry of the order
parameter, and orthorhombic distortions of the lattice are also studied. 
\end{abstract}
\end{minipage}}

\maketitle
\date{\today}

\section*{Introduction}
Kamerlingh Onnes discovered some ninety years ago that the resistivity of 
certain metals would drop to zero when the temperature was lowered below a 
critical value $T_c$. He accordingly termed superconductors all materials which 
exhibited such state of disipationless electric flow. Superconductors display 
as a matter of fact a rich variety of physical phenomena other than
superflow, some of which are closely related to it, like the Meissner effect,
some others having a different nature, as the electronic gap which appears
in tunneling experiments. 

Superflow and Meissner's effects are a consequence of the London equation, a 
linear relation between the density current and the vector potential, 
$\vec{\j}+{\sf K}\, \vec{A}=0$, which serves to define the current response
tensor ${\sf K}$. Such an equation, or its non-local version, holds 
in a superconductor even in the presence of the usual scattering mechanisms 
which degrade the current in a normal metal. A conventional exercise consists in 
solving London and Maxwell's
equations in a specific geometry to find the distribution of current densities 
and magnetic fields. \cite{reitz_book,pyun89a} It is shown in
this way that they are highly nonuniform across the sample: they are 
confined to a thin shell in contact to the surface and do not penetrate the 
bulk of the sample.

Interestingly enough, superflow and Meissner's effects might be the most 
difficult properties to extract from the 
microscopic BCS theory of superconductivity, because the derivation of the
London equation is not an easy task. The standard procedure to derive this
relation uses linear response theory. \cite{schrieffer_book}
One complication arises from the fact that such expression is not gauge 
invariant, so that a change in the longitudinal part of $\vec{A}$ will change 
$\vec{\j}_l$. The longitudinal
response kernel ${\sf K}_l$ is in turn related to the density response
function through the continuity equation, so the problem of gauge invariance
can also be stated as a problem of charge conservation.
Within linear response theory, BCS theory must therefore be extended to enforce 
the Ward Identity which relates the dressed vertex of the longitudinal response
function ${\sf K}_l$ to the BCS self-energy.\cite{schrieffer_book}  It can
also be proven that any explicit computation of superflow in a superconductor 
must be fully self-consistent in order to ensure that charge is 
conserved.\cite{ferrer_thesis,furusaki91a}

The superflow state can be destroyed by increasing the current 
density, the magnetic field, or the temperature above certain critical values 
which depend among other things on the specific material used, the amount of 
disorder of the sample, or the geometry of the device which is used. 
Furthermore, 
the relation between $\vec{\j}$ and $\vec{A}$ is not linear for large 
$\vec{\j}$. In other
words, London's equation does not hold close to the critical value of the 
superflow current density, ${\rm j}_c$. The situation in strong type II 
superconductors is 
even more complicated because the magnetic field originated by the supercurrent 
might give rise to vortices; these vortices move in the
presence of the supercurrent and dissipate energy.\cite{kim_parks_book} 

Superflow in High-temperature  superconductors might be anisotropic because 
(a) the order parameter has $d$-wave symmetry; (b) the electronic 
structure is also fairly anisotropic, with van Hove 
singularities slightly below the Fermi energy;\cite{dessau93a,gofron94a} (c) the
two-dimensional ${\sf Cu\,O_2}$ planes, responsible for much of the physics of
the cuprates, present different types of orthorhombic distortions. 
The purpose of this article is to study the physics of superflow in clean,
infinite, $d$-wave superconductors on an orthorhombically distorted square 
lattice.
A phenomenological two-dimensional tight-binding model which captures some of
the experimental features of the superconducting state of High-temperature
superconductors will be used.\cite{norman95a,ferrer98a} No attention will be 
paid to the effects of disorder, twin boundaries, or external or induced 
magnetic
fields, all of which can be important in this compounds and will be dealt with
in the near future. Superflow in two-dimensional $d$-wave superconductors has 
already been studied using a Ginsburg-Landau approach, which focused in 
particular on the induction of non-$d$-wave components upon switching on of the 
current.\cite{zapotocky97a} Superflow in quasi-one-dimensional $s$-wave 
superconductors has also been studied in
equilibrium\cite{maki_parks_book,bagwell94a} and non-equilibrium 
situations.\cite{canizares97a,martin95a}

Section II is devoted to present the model and the formalism used to solve it in
the presence of a finite superflow. The mapping onto a linearized isotropic 
model is also shown. Section III discusses the results obtained for the phase 
diagram of the model, the gaps, their distortion due to orthorhombic effects 
and the local density of  states, for zero current density. 
Section IV is devoted to study the doping and angular dependence of the gaps,
free energy, current density, and critical current in the presence of a
superflow. A summary closes the
article. Even though the analytical calculations have been performed
using the Matsubara formalism, all the numerical results provided 
in the article correspond to T=0.

\section*{Formalism}
The partition function of a superconducting sample in the presence of an 
electromagnetic field can be written as  

\begin{equation}
{\cal Z}(T,\mu)=\int {\cal D}[\Delta^*,\Delta,
\vec{A},A_0]\, e^{-\frac{1}{\hbar}\,{\cal S}
[\Delta^*,\Delta,\vec{A},A_0]}
\end{equation}
where the functional integral sums over all possible configurations of order
parameter and electromagnetic fields, 
$\Delta(x,x')=|\Delta(x,x')|\,e^{i\,(\phi(x)+\phi(x'))},\,\vec{A(x)},\,A_0(x)$, 
and where the imaginary-time action is
\begin{eqnarray}
{\cal S}
&=&\int dx \, dx' \Bigl[-{\rm Tr} \,\ln\,\tilde{G}_{\sf e}^{-1}(x,x')+
\frac{|\Delta(x,x')|^2}{V(x,x')}\nonumber\\&&\\&&+\Bigl(e\,n_i
\,A_0(x)+\frac{\vec{B}^2(x)-\vec{E}^2(x)}{8\,\pi}\Bigr)\,\delta(x-x')
\Bigr]+{\cal H}^*(0)\nonumber.
\end{eqnarray}

\noindent Here $x$ denotes the four vector $(\tau,\vec{x})$, 
$V(x,x')= V(\vec{x}-\vec{x'})\,\delta(\tau-\tau')$ is the pairing 
potential, $e\,n_i$ is the background charge density of ions,
$\tilde{G}_{\sf e}(x,x')$ 
is the Green function of electrons in Nambu's space,
\begin{eqnarray}
\left( \begin{array}{cc}
\Bigl(\hbar \,\partial_{\tau} +{\cal H}(x)\Bigr)\,\delta(x-x')&
\Delta(x,x')\,\delta(\tau-\tau')\\
\Delta^*(x,x')\,\delta(\tau-\tau')&
\Bigl(\hbar \,\partial_{\tau} -{\cal H}^*(x)\Bigr)\,\delta(x-x')
\end{array}\right)^{-1},\nonumber
\end{eqnarray}
and in the Hamiltonian
\begin{displaymath}
{\cal H}(x)=\frac{(\vec{p}+e/c\,\vec{A}(x))^2}{2\,m}+
v_{latt}(\vec{x})-e\,A_0(x)-\mu
\end{displaymath}
the Zeeman term has been discarded.

It can be seen that the action is invariant under the following gauge 
U(1) symmetry
\begin{eqnarray}
A_0(x)&\mapsto& A_0(x)-\frac{i\,\hbar}{2\,e}\partial_\tau \theta(x),\nonumber\\
\vec{A}(x)&\mapsto& \vec{A}(x)+\frac{\hbar\,c}{2\,e}\vec{\nabla}\theta(x),\\
\phi(x)&\mapsto& \phi(x)-\frac{\theta(x)}{2},\nonumber
\end{eqnarray}
up to a total time derivative.

The phase of the order parameter field can therefore be absorbed by the 
electromagnetic field, and drops out of the problem if one defines the gauge 
invariant quantities\cite{otterlo97a}
\begin{eqnarray}
\label{superfluid1}
&&\Psi_s(x)=A_0(x)-\frac{i\,\hbar}{2\,e}\,\partial_\tau\,\phi(x),\nonumber\\
&&\vec{v}_s(x)=\frac{\hbar}{2\,m}\left(\vec{\nabla}\phi(x)+\frac{2\,e}{\hbar\,c}
\,\vec{A}(x)\right),
\end{eqnarray}
in terms of which the electric and magnetic fields are written as
\begin{eqnarray}
&&\vec{E}(x)=-\vec{\nabla}\,\Psi_s(x)-\frac{i\,m}{e}\,\partial_\tau\,
\vec{v}_s(x),\nonumber\\
&&\vec{B}(x)=\frac{m\,c}{e}\,\vec{\nabla}\times\vec{v}_s(x).\label{superfluid2}
\end{eqnarray}

As shown by van Otterlo and coworkers,\cite{otterlo97a} the whole problem 
can be rephrased using these new variables:
\begin{eqnarray}
&&{\cal Z}(T,\mu)=\int {\cal D}[\Delta_l,
\vec{v}_s,\Psi_s]\, e^{-\frac{1}{\hbar}\,{\cal S'}
[\Delta_l,\vec{v}_s,\Psi_s]},\nonumber\\\nonumber\\
&&
{\cal S'}
=\int dx \, dx' \Bigl[-{\rm Tr} \,\ln\,\tilde{G}_{\sf e}^{-1}(x,x')+
\frac{\Delta_l^2(x,x')}{V(x,x')}\nonumber\\&&+\Bigl(e\,n_i
\,\Psi_s(x)+\frac{\vec{B}^2(x)-\vec{E}^2(x)}{8\,\pi}\Bigr)\,\delta(x-x')
\Bigr]+{\cal H}^*(0)\nonumber,\\\nonumber\\
&&\tilde{G}_{\sf e}^{-1}=
\left( \begin{array}{cc}
\Bigl(\hbar \,\partial_{\tau} +{\cal H}\Bigr)\,\delta(x-x')&
\Delta_l\,\delta(\tau-\tau')\\
\Delta_l\,\delta(\tau-\tau')&
\Bigl(\hbar \,\partial_{\tau} -{\cal H}^*\Bigr)\,\delta(x-x')
\end{array}\right),\nonumber\\\nonumber\\
&&{\cal H}(\vec{x})=\frac{(\vec{p}+m\,\vec{v}_s(x))^2}{2\,m}+
v_{latt}(\vec{x})-e\,\Psi_s(x)-\mu,\label{superfluid3}
\end{eqnarray}
where $\Delta_l=|\Delta|$ is a real scalar field whose vacuum value will be 
different from zero below $T_C$ and the phase $\phi$ is explicitly linked 
to the dynamics of the electromagnetic fields and the Higgs mechanism through 
the longitudinal part of the superfluid velocity field, $\vec{v}_{s,l}$.

Stationarity of the action with respect to $\Delta_l, \vec{v}_s$ and $\Psi_s$
leads to the saddle point equations which determine the classical configurations
of electromagnetic and order parameter fields, and to the saddle point
approximation for the thermodynamic potential
\begin{mathletters}
\begin{equation}
\vec{\nabla}\cdot\,\vec{E}_{cl}(x)=4\,\pi\,e (n_i-n_e(x)),\label{SPa}
\end{equation}
\begin{equation}
-\frac{i}{c}\partial_\tau\vec{E}_{cl}(x)+\vec{\nabla}\times\vec{B}_{cl}(x)=
\frac{4\,\pi}{c}{\sf \vec{\j}}(x),\label{SPb}
\end{equation}
\begin{equation}
\Delta_{l,cl}(\vec{x},\vec{x'})=-\frac{1}{L^d}\,V(\vec{x},\vec{x'})\,
G_{{\sf e},12}(\vec{x},\vec{x'}),\label{SPc}
\end{equation}
\begin{equation}
\Omega_{SP}(T,\mu,\Delta_{l,cl},\vec{E}_{cl},\vec{B}_{cl})= 
\frac{1}{\beta\,\hbar}{\cal S}_{SP}'
[\Delta_{l,cl},\vec{E}_{cl},\vec{B}_{cl}],\label{SPd}
\end{equation}
\end{mathletters}
where $L^d$ is the volume of the sample, and $G_{{\sf e},12}$ is the
off-diagonal element of $\tilde{G}_{\sf e}$. The electronic and current 
densities are defined in terms of the diagonal elements of $G_{\sf e}$ as
\begin{mathletters}
\begin{eqnarray}
&&n_e(x)=\frac{G_{{\sf e},11}(x,x)-G_{{\sf e},22}(x,x)}{L^d},\label{dens}\\
&&{\vec {\j}}_e(x)=-\frac{e}{m\,L^d}\,Tr
\,\left[\left(\vec{p}\,\tilde{\openone}+m\vec{v}_s(x)\,
\tilde{\tau}_z\right)\,\tilde{G}_{\sf e}(x,y)\right]_{y=x},\label{curr}
\end{eqnarray}
\end{mathletters}
where $\tilde{\tau}_z$ is the third Pauli matrix.

The saddle point equation for the order parameter can be further simplified
by changing to center of mass and relative coordinates and performing the
Fourier transform with respect to the relative coordinate. Then
\begin{equation}
\Delta_{l,cl}(\vec{k},\vec{R})=-\frac{1}{L^d}\sum_{\vec{k'}} \, 
V(\vec{k}-\vec{k'})\, G_{e,12}(\vec{k'},\vec{R}).
\end{equation}

Expanding now $\Delta_{l,cl}$ and $V$ in terms of some basis set of functions
$\eta_a$ as can be, for instance, those which form the irreducible 
representations of a crystallographic point group,
\begin{eqnarray}
&&\Delta_{l,cl}(\vec{k},\vec{R})=\sum_a
\Delta_{a}(\vec{R})\,\eta_a(\vec{k}),\nonumber\\
&&V(\vec{k}-\vec{k'},\vec{R})=\sum_a V_a(\vec{R})\,\eta_a(\vec{k})\,
\eta(\vec{k'})
\end{eqnarray}
one easily finds that
\begin{equation}
\Delta_a(\vec{R})=-\frac{V_a(\vec{R})}{L^d}
\sum_{\vec{k}}\,\eta_a(\vec{k})\,G_{e,12}(\vec{k},\vec{R}).
\end{equation}

Superflow is a superconducting state characterized by the stationary flow of a
current $\vec{\j}$ in the absence of any longitudinal electric field (neglecting
that built up at the boundary of the specimen.) This 
implies in view of Eqs.\ (\ref{superfluid1}) and (\ref{superfluid2}) that 
$\Psi_s(x)$ is 
identically zero, and that the superfluid velocity $v_s(x)$ does not depend on 
time. This state is therefore determined by two classical fields,
$\Delta_{l,cl}(\vec{x})$ and $\vec{v}_{s,cl}(\vec{x})$, which must be determined 
self-consistently with the aid of Eqs.\ (\ref{SPb}) and (\ref{SPc}). 

In an infinite sample, with no externally applied fields, $\Delta_{l,cl}$ and
$\vec{v}_{s,cl}$ are uniform, so that Eq.\ (\ref{SPb}) drops out of the problem 
and 
the superfluid velocity is just an input parameter, which can be viewed as a
Doppler shift acquired by the kinetic energy part of the Hamiltonian
\begin{equation}
{\cal T}=\frac{(\vec{p}+m\,\vec{v}_s)^2}{2\,m}
=\frac{(\vec{p}+\hbar\,\vec{q})^2}{2\,m},
\end{equation}

The order parameter field must still be determined and, once self-consistency is
achieved, the current density can be computed via Eq.\ (\ref{curr}). 
The continuity equation, $\vec{\nabla}\cdot \vec{\j}=0$ is automatically
satisfied in this case.
  
In a finite geometry with no constrictions, like a wire or a thin film, the full
self-consistent problem must be handled, but there are some facts which simplify 
the computation: (a) $\vec{\j}(\vec{x})$ must always point in the same 
direction, say the Z axis, because of the symmetry of the problem. (b) The 
continuity equation then demands that $\vec{\j}$ and $\vec{v}_s$ vary only in 
the XY plane. Now, the continuity equation is automatically satisfied if the
order parameter is calculated self-consistently.\cite{ferrer_thesis,furusaki91a}
Therefore, any explicit self-consistent calculation conserves charge and is 
gauge invariant by construction if $\vec{\j}$ is computed using 
Eq.\ (\ref{curr})
with no further approximations like linear response theory.  
     
For an infinite superconductor, the current density can also be obtained in a 
more direct fashion by calculating the gradient of the thermodynamic potential,
\begin{equation}
\vec{\j}[q]=-\frac{e}{\hbar \,L^{d-1}}\,\vec{\nabla}_{\vec{q}} \,
\Omega_{SP}(T,\mu,[q]).
\end{equation}

It is worth pointing out that
a finite Doppler shift in the dispersion relation of a normal metal leads to no
observable effects, because it can always be made disappear by performing a 
simple change of variables in the k-sums, $\vec{k}+\vec{q}\mapsto \vec{k}$.
Indeed, a finite current can only be induced in a metal by applying a finite 
electric field.

The formalism developed so far is going to be applied to study the physics 
of superflow in a clean, infinite $d$-wave superconductor where the 
electromagnetic fields are zero or at least negligible. The electronic fields 
will be described by a phenomenological one-band Hubbard-like model on an 
orthorhombically distorted square lattice. \cite{norman95a,ferrer98a}

The dispersion relation used comes from a fit to the 
spectral peaks found in ARPES experiments of optimally doped
${\sf Bi_2 Sr_2 Ca Cu_2 O_{8-\delta}}$. \cite{norman95a} The interactions among
electrons $V(\vec{x}-\vec{x'})$ are assumed to be different from zero only when
$\vec{x}$ and $\vec{x'}$ are the same site ($V(\vec{0})=V_0$) or when they are 
nearest neighbors ($V(\vec{x}-\vec{x'})=V(\vec{\delta})= V_1$). Explicitly,
  
\begin{eqnarray}
{\cal H}= &&\sum_{\vec{k},\sigma}\left[\frac{1}{z}\sum_{\vec{\delta}} 
\,t(|\vec{\delta}|)\, e^{i\,\vec{k}\cdot\vec{\delta}}\,-\mu\right] 
\hat{c}_{\vec{k},\sigma}^+ \,\hat{c}_{\vec{k},\sigma}\nonumber\\
&&-\,V_0 \,\sum_i \: \hat{c}_{i,\uparrow}^+ \,\hat{c}_{i,\downarrow}^+
\,\hat{c}_{i,\downarrow} \,\hat{c}_{i,\uparrow}
\,-\,V_1 \,\sum_{i,\delta'} \: \hat{c}_{i,\uparrow}^+\, 
\hat{c}_{i+\delta',\downarrow}^+\,
\hat{c}_{i+\delta',\downarrow} \,\hat{c}_{i,\uparrow},\nonumber\\&&
\end{eqnarray}
where $z$ is the coordination number, the sum in $\vec{\delta}$ extends up to 
five nearest neighbors, the 
hopping integrals $t(|\vec{\delta}|)$ have the values [-595,164,-52,56,51] 
meV, and the chemical potential can be varied freely 
to change the mean occupation number $n$ of the band. It is therefore 
assumed that the band is shifted rigidly upon doping, a fact which is not 
inconsistent with the available data from ARPES experiments. \cite{ding97a}

Monte Carlo and RPA studies of the repulsive Hubbard model have found that the
effective potential among dressed quasiparticles assumes an RKKY or "Mexican
hat" form \cite{scalapino95a} so that it has a large repulsive on-site 
contribution and then decays exponentially in an oscillating way. Within this 
magnetic scenario for superconductivity in the cuprates, \cite{chubukov96a}
the length scale over which the potential varies is set by the magnetic 
coherence length $\xi_{mag}$ and is therefore uncommensurated with the lattice
spacing $a$. The interaction term used in this article bears some resemblance 
with such an effective potential if the on-site interaction $V_0$ if assumed to 
be large and negative. The resulting effective model can then be thought of as 
a low energy fixed point 
Hamiltonian which appears after integrating out the high frequency degrees of 
freedom of a more complicated model describing the physics of the cuprates. The 
effective model can therefore be safely solved using Mean Field techniques. 

Generically, the four possible one-dimensional irreducible representations of 
the group of the square $C_{4v}$ transform under its symmetry operations as the 
identity, $x^2-y^2$, $x\,y$ and $x\,y\,(x^2-y^2)$.
\cite{annett_ginsbergV_book} Their explicit form in a tight-binding formulation
would be $s$ or extended $s$ ($s^*$), $d_{x^2-y^2}$, $d_{xy}$, and  
$d_{xy}\times d_{x^2-y^2}$, states.

A Mean 
Field decomposition of the on-site interaction naturally gives rise to a pure 
$s$ or to a spin-density-wave (SDW) state, depending on whether $V_0$ is 
attractive or repulsive. A nearest neighbor attraction produces an $s^*$ and a 
$d_{x^2-y^2}$ state; a second nearest neighbors interaction $V_2$ gives rise to 
a second $s^*$ state and to the $d_{xy}$ state, and so on. 
The Mean Field solution of this Hamiltonian therefore allows for a 
SDW state, and superconducting $s$-wave, extended $s$-wave, $d$-wave states, 
and mixed 
representations of them ($s+s^*,d+i(s+s^*),s+s^*+d$). Similar results have been
obtained in Refs.\ \onlinecite{micnas90a,nazarenko96a,feder97a,bealmonod96a}. 

The saddle point equations for the thermodynamic potential, the order parameter,
the occupation of the band and the Helmholtz free energy per site are
\begin{mathletters}
\begin{eqnarray}
\frac{\Omega_{SP}(T,\mu)}{M}=&&\frac{|\Delta_s|^2}{V_0}+ 
\frac{|\Delta_{s^*}|^2+|\Delta_d|^2}{V_1}\\&&+\frac{1}{M}
\sum_{\vec{k}} \left[\xi_{\vec{k}+\vec{q}}-E_{\vec{k},\vec{q}}-\frac{2}{\beta} 
\ln (1+e^{-\beta E_{\vec{k},\vec{q}}}) \right]\nonumber, 
\end{eqnarray}
\begin{equation}
\Delta_a=\frac{V_{a}}{M} \sum_{\vec{k}} \frac{\Delta_{\vec{k}} 
\,\eta_{a,\vec{k}}}{2 E_{\vec{k},\vec{q}}}
\tanh\left(\frac{\beta E_{\vec{k},\vec{q}}}{2}\right),
\label{SP_TB}
\end{equation}
\begin{equation}
n=\frac{N}{M}=\sum_{\vec{k}}\left[1-
\frac{\xi_{\vec{k}+\vec{q}}+\xi_{\vec{k}-\vec{q}}}{2\,E_{\vec{k},\vec{q}}} 
\tanh\left(\frac{\beta E_{\vec{k},\vec{q}}}{2}\right)\right],
\end{equation}
\begin{equation}
f_{SP}(T,n)=\frac{\Omega_{BCS}(T,\mu(n))}{M}+\mu(n)\,n,
\end{equation}
\end{mathletters}
where the dispersion relation for the quasiparticles is 
\begin{eqnarray}
E_{\vec{k},\vec{q}}&=&\frac{\xi_{\vec{k}+\vec{q}}-\xi_{\vec{k}-\vec{q}}}{2}+
\sqrt{\left(\frac{\xi_{\vec{k}+\vec{q}}+\xi_{\vec{k}-\vec{q}}}{2}\right)^2+
|\Delta_{\vec{k}}|^2},\nonumber\\
\Delta_{\vec{k}}&=&\sum_{a=s,s^*,d} \Delta_a \eta_{a,\vec{k}},
\end{eqnarray}
and
the $\eta$-functions, which form a truncated tight-binding basis of the crystal point 
group $C_{4v}$, are given in table I, along 
with the allowed range of variation of $V_0$ and $V_1$. It has been assumed in
the numerical calculations of this article that $\xi$ is equal to the lattice
spacing $a$, but it can generically be any other characteristic length scale,
like the magnetic coherence length $\xi_{mag}$.   

\begin{table}
\caption{Range of variation of the interactions along with their
respective order parameters and $\eta$-functions}
\label{table1}

\begin{tabular}{cccc}
Parameters & Range (meV) &Gaps&$\eta_{a,\vec{k}}$  \\\hline
$V_0$     &  [-400,400]    & $\Delta_s$      &  $1$                        \\
$V_1$     &   [0,1500]     & $\Delta_{s*}$   & $cos(k_x \xi)\,+\,cos(k_y \xi)$   \\
$V_1$      &  [0,1500]     & $\Delta_d$      & $cos(k_x \xi)\,-\,cos(k_y \xi)$   \\
\end{tabular}
\end{table}

Orthorhombic effects are present in most of the cuprates, giving rise to a 
mixing of the irreducible representations. For 
instance, in ${\sf Y Ba_2 Cu_3 O_{8-\delta}}$, the x and y axes become 
inequivalent, so that the $s$ and $d_{x^2-y^2}$ states on the one hand and the 
$d_{xy}\times d_{x^2-y^2}$ and $d_{xy}$ states on the other are mixed. For 
${\sf La_{2-x}\,Sr_x\,Cu\,O_4}$ and ${\sf Bi_2 Sr_2 Ca Cu_2 O_{8-\delta}}$, it 
is the two orthogonal $45^0$ axes which become inequivalent and then the mixing 
pairs are $s$ and $d_{xy}$ on the one hand and $d_{x^2-y^2}$ and 
$d_{xy}\times d_{x^2-y^2}$ on the other. \cite{annett_ginsbergV_book} 

Orthorhombic distortions on the square lattice should affect both the hopping
integrals and the nearest neighbor interactions. These effects can be taken 
into account by letting the hopping amplitudes $t_i$ and the nearest neighbor 
interaction $V_1$ be different along the X and Y axes, as is the case of 
${\sf Y Ba_2 Cu_3 O_{8-\delta}}$, or the $45^0$ axes (this would correspond to 
${\sf La_{2-x}\,Sr_x\,Cu\,O_4}$ and ${\sf Bi_2 Sr_2 Ca Cu_2 O_{8-\delta}}$). 
For example, for ${\sf Y Ba_2 Cu_3 O_{8-\delta}}$, the hopping integrals 
$t_1,\,t_3,\,t_4$ change to $t_{i,(x,y)}=t_i\,(1\pm\alpha)$ and 
$V_{1,(x,y)}=V_1\,(1\pm\beta)$.

The expression for the current density
\begin{eqnarray}
{\cal \vec{\j}}=&&-\frac{2\,e}{\hbar\,a}\int_{\pi}^{\pi}\,
\frac{d\vec{k}}{(2\,\pi)^2}\,\left[\vec{\nabla}_{\vec{q}}
\,\xi_{\vec{k}+\vec{q}}\, |v_{\vec{k}}|^2\right.\nonumber\\&&+\left.
\left(\vec{\nabla}_{\vec{q}}\,\xi_{\vec{k}+\vec{q}}\,|u_{\vec{k}}|^2-
\vec{\nabla}_{\vec{q}}\,\xi_{\vec{k}-\vec{q}}\,|v_{\vec{k}}|^2\right)
n_F(E_{\vec{k},\vec{q}})\right],\label{current}
\end{eqnarray}
has two terms. The main contribution comes from the condensate; 
the other is a back-flow term which comes from the quasiparticles and partially 
counteracts the contribution from the condensate. The quasi-particle term is also
responsible for the non linear behavior at large values of $\vec{q}$. The
lattice spacing $a$ has been set equal to the experimental in-plane lattice 
constant of
${\sf Bi_2 Sr_2 Ca Cu_2 O_{8-\delta}}$, which is 5.4 $\AA$.  
 
The above expression for $\vec{\j}$ is valid to all orders in $\vec{q}$ and
therefore goes well beyond linear response theory. To find the superfluid
density tensor, ${\vec{\j}}$ should be expanded to linear order in $\vec{q}$ 
\begin{eqnarray}
{\j}_i&=&-\frac{2\,e}{\hbar\,a}\int_{\pi}^{\pi}\,\frac{d\vec{k}}{(2\,\pi)^2}\,
|v_{\vec{k}}|^2\tanh \left(\frac{\beta\,E_{\vec{k}}}{2}\right)\,
\frac{\partial^2\,\xi_{\vec{k}}}{\partial k_i\,\partial k_j}\;q_j\nonumber\\
&=&-2\,e\,m\,\int_{\pi}^{\pi}\,\frac{d\vec{k}}{(2\,\pi)^2}\,
|v_{\vec{k}}|^2\tanh \left(\frac{\beta\,E_{\vec{k}}}{2}\right)\,
\left[M^{-1}\right]_{ij}\,v_{j,s}\nonumber\\&=&-2\,e\,n_{s,i,j}\,v_{j,s},
\end{eqnarray}
so that
\begin{equation}
n_{s,i,j}=m\,\int_{\pi}^{\pi}\,\frac{d\vec{k}}{(2\,\pi)^2}\,
|v_{\vec{k}}|^2\tanh \left(\frac{\beta\,E_{\vec{k}}}{2}\right)\,
\left[M^{-1}\right]_{ij}.
\end{equation}
Here $\left[M^{-1}\right]$ denotes the mass tensor. The current is therefore
not parallel to the superfluid velocity, although the angle formed by them is
always small. 

A mapping can be done from the tight-binding Hamiltonian onto a much simpler
linearized model,\cite{musaelian95a} by expanding all functions and keeping 
only terms linear or quadratic in $\vec{k}$. The following substitutions must 
then be made
\begin{eqnarray}
&&\xi_{\vec{k}}\:\mapsto\: 2 \:t_R \:k_F \lambda\:(k-k_F) \lambda
=\hbar \,v_F \,(k-k_F) \:\mapsto\: \varepsilon,\nonumber\\\nonumber\\
&&\int_{-\pi/a}^{\pi/a}\frac{d\vec{k}}{(2\pi)^2}
\:\mapsto\:{\rm g}^L\,\int_0^{2\pi}
\frac{d\theta}{2\pi}\int_0^{E_d}d\varepsilon ,\nonumber\\\nonumber\\
&&\Delta_{\vec{k}}\:\mapsto\:
\Delta^L(\theta)=\Delta_s^L\,\eta_s^L(\theta)+\Delta_d^L \, \eta_d^L(\theta)=
\Delta_s^L+\Delta_d^L \,\cos(2\theta).\nonumber\\
\end{eqnarray}
where $\lambda$ is some length scale of the order of the lattice constant.

The saddle point equation for the order parameters fields can be written now
as
\begin{eqnarray}
\Delta_{s,d}^L=&&{\rm g}^L \,V_{s,d}^L \,\int_0^{2\pi}\frac{d\theta}{2\pi}
\int_{0}^{E_D} d\varepsilon\,
\frac{\eta_{s,d}^L(\theta)\, \Delta^L(\theta)}
{\sqrt{\varepsilon^2+|\Delta^L(\theta)|^2}}\nonumber\\
&&\times\tanh\left(\frac{\beta \,
(\hbar \vec{v}_F\cdot \vec{q}
+\sqrt{\varepsilon^2+|\Delta^L(\theta)|^2})}{2}\right),
\end{eqnarray}
where
\begin{displaymath}
\begin{array}{ll}\\
{\rm g}^L\, V_s^L= {\rm g}\, V_0,&\:\:\:\Delta_s^L=\Delta_s,\\
{\rm g}^L\, V_d^L= {\rm g}\,\gamma^2\,V_1,&\:\:\:\Delta_d^L=\gamma\,\Delta_d,\\\\
\gamma=-\frac{(k_F\lambda)^2}{2}+\frac{(k_F\lambda)^4}{24},&\:\:\:
{\rm g}=\frac{1}{4\,\pi\,t_R}.
\end{array}
\end{displaymath}

There are three parameters, namely $k_F\,\lambda$, g, and $E_d$, which 
can be adjusted to fit the values of $\Delta_d$ and $\Delta_s$ along the $V_0$ 
and $V_1$ axes. 

\section*{Discussion of zero current properties}

\begin{figure}
\psfigurepath{plots}
\psfig{figure=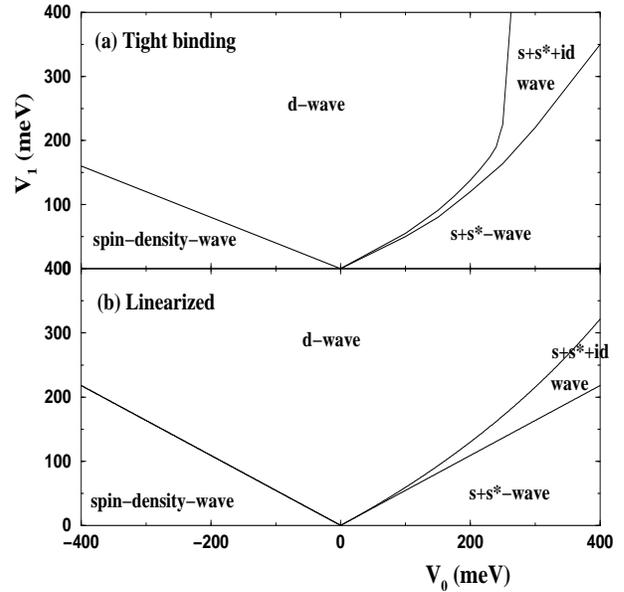,height=8cm,width=8cm,angle=270}
\vspace{0.5cm}
\caption{Phase diagram for (a) the tight-binding model with doping x=0.2; (b)
the linearized model, with parameters chosen to fit the order parameter fields
along the axes.}
\label{fig1}
\end{figure}

The numerical solution of the saddle points equations \ (\ref{SP_TB}) for the 
order parameters is pretty straightforward. The zero temperature phase diagram
is a function of the two coupling constants $V_0$ and $V_1$, and the occupation
number $n$ or, equivalently, the doping concentration $x=1-n$. 
Fig.\ 1 (a) shows a representative case, taken at $x=0.2$. There exist
SDW, pure superconducting $s$ and $d$, and mixed $s+s^*$, $s+s^*+d$, and 
$d+i(s+s^*)$ states.\cite{odonovan95a}
The $d$-wave
solution exists and is more stable close to half-filling, for values of $V_1$
larger than $|V_0|$. The s-state solutions, on the contrary are more stable for
small occupations of the band or when $V_1<V_0$. The time-reversal-symmetry
broken solution $d+i(s+s^*)$ exists and is always more stable along the phase 
boundary between the d and the mixed $s+s^*$ states. The mixed $s+s^*+d$ only
exists for large values of $V_1$ and therefore doesn't appear in Fig.\ 1 (a). 

The effective on-site interaction in the case of the cuprates must still be 
repulsive and large even after a hypothetical process of renormalization of high 
frequency degrees of freedom. The part of the phase diagram relevant to these 
compounds is therefore the negative $V_0$ quadrant, where the only physical 
phases are $d$-wave and SDW states. This model then nicely produces 
competition between $d$-wave superconductivity and antiferromagnetism, 
even though their phase boundary occurs for ratios $V_1/|V_0|$ which are too 
large as compared with the "Mexican hat" effective potential referred to in the 
previous section. The model also 
excludes any exotic time-reversal-symmetry broken state, like the $d+i(s+s^*)$.
Quantum fluctuations will likely melt the ordered phases into a strongly 
correlated electron liquid in the neighborhood of their phase boundary as in 
other correlated low-dimensional systems with competing 
states.\cite{chandra88a,ferrer93a} Indeed, they could also displace the 
phase boundary towards larger values of $|V_0|$. 
The numerical results presented for the tight-binding model
in the rest of the paper will use the values $(V_0,V_1)=(-150,250)$ meV for the
interaction constants, which correspond to a superconducting solution in the 
neighborhood of the phase boundary between $d$-wave and SDW  states. 

\begin{figure}
\psfigurepath{plots}
\psfig{figure=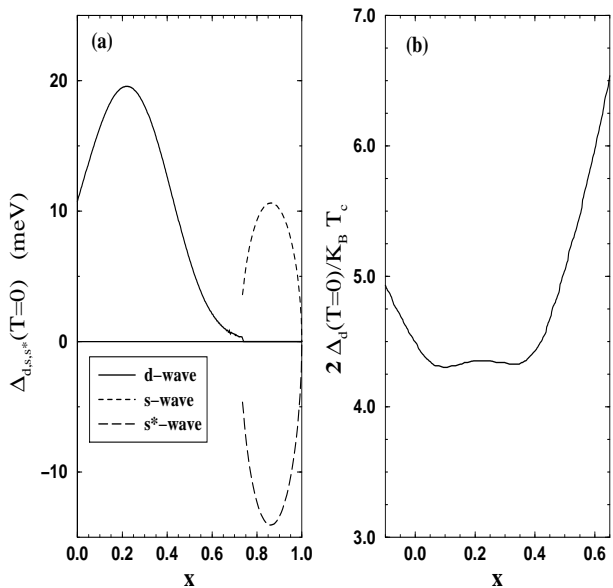,height=8cm,width=8cm,angle=270}
\vspace{0.5cm}
\caption{(a) Gaps $\Delta_{d,s,s^*}$; (b) ratio 
$2\Delta_d/K_B T_c$ as a function of doping x.}  
\label{fig2}
\end{figure}

The doping dependence of the zero temperature gaps is plotted in Fig.\ 2 (a). 
The amplitude of the $d$-wave gap is always larger 
close at half-filling and vanishes at about x=0.7. The maximum occurs at x=0.22 
irrespective of the values of the coupling constants due to fact that the Fermi 
level crosses over a van Hove singularity. Interestingly enough, there are 
$s$-wave solutions of the saddle point equations for large doping concentrations 
even though the on site interaction is repulsive. The $s$-wave gaps peak at 
$x\simeq 0.8$ because the Fermi level crosses a second van Hove singularity. 

The ratio $2\,\Delta_d/K_B\,T_c$ (Fig.\ 2 (b)) has a constant value of 4.3 for 
doping concentrations ranging from x=0.05 to 0.4 (the ratio for a parabolic 
band is universal and equal to 4.25.) The highest value of $T_c$ therefore 
occurs at x=0.22, which is defined at the optimal doping concentration of this 
model. The ratio increases steeply for larger x, reaching a value of 7 when x 
is 0.67. Hence, the doping dependence of the critical temperature resembles the 
universal curve $T_c(x)$ of the cuprates.\cite{feder97a,duffy97a,spathis92a} 
Moreover, all physical magnitudes are expressed in real units, so that one may 
compare with experiments easily. The agreement doesn't extend too far though, 
as the experimental doping dependence of the zero temperature gap is a 
monotonically decreasing function of x. \cite{ding97a,miyakawa98a} 
The experimental ratio $2\Delta_d(T=0)/K_B\,T_c$ is therefore non constant.

\begin{figure}
\psfigurepath{plots}
\psfig{figure=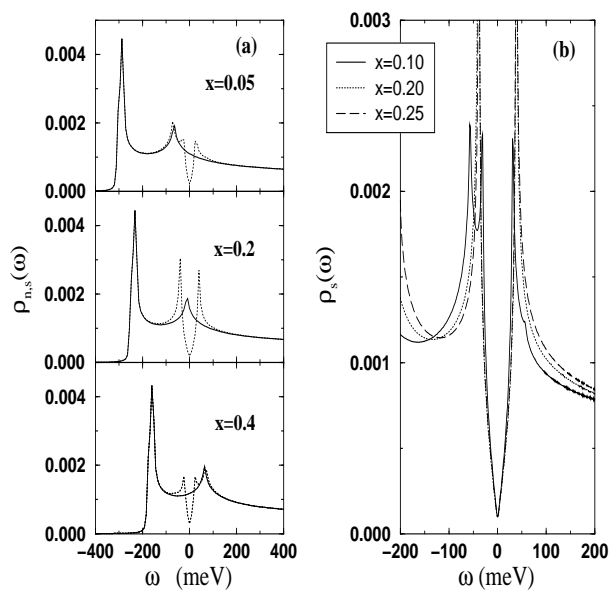,height=8cm,width=8cm,angle=270}
\vspace{0.5cm}
\caption{(a) Normal $\rho_n(\omega)$ and superconducting $\rho_s(\omega)$ 
densities of states for three doping
concentrations in a frequency range which covers almost all the band; 
(b) superconducting density of states $\rho_s(\omega)$ for various x in the 
narrow frequency range [-200,200] meV. }
\label{fig3}
\end{figure}

For underdoped cuprates there appears another characteristic temperature 
$T^*$, which is a linearly decreasing function of the doping concentration. 
Above $T_c$ but below $T^*$, 
the $d$-wave gap partially closes down in such a way that the Fermi contour 
exists
only in segments around the diagonals of the two-dimensional Brillouin zone.
\cite{ding97a} 
Assuming that both $T^*$ and $\Delta_d(T=0)$ vanish at a doping concentration 
of about 0.3, the dependence of both magnitudes with x is close to a straight 
line of slope $p_{T,\Delta}$. A possible generalization of $2 \Delta/K_B\,T_c$ 
for the cuprates could therefore be the ratio $2*p_{\Delta}/p_T$, which is 
approximately constant and equal to about 5.5.  

The main source of trouble probably comes from the simple form of the dispersion 
relation of the model, where several drastic assumptions about the existence and
nature of the hypothetical quasiparticles have been made: (a) There exists a
well defined Fermi surface, even though this system has a pseudo gap above 
$T_c$. (b) Quasiparticles not 
only exist and are infinitely long-lived, but also their weight on the 
renormalized electronic propagator is always exactly 1. (c) The band
shifts rigidly when the system is doped. To obtain the correct behavior for the 
doping dependence of $\Delta_d(T=0)$, the treatment of the dispersion
relation of the fermionic degrees of freedom should be improved. In particular,
if one still wishes to use a quasi-particle picture, at least assumption (b) 
should be relaxed.   
  
The specific dispersion relation used gives rise to a second, weak van Hove 
singularity, in addition to that proper of the Hubbard t-t' model, see Fig.\ 
3 (a). They are responsible for the maximums attained by the zero temperature 
gaps as a function of doping. The dispersion relation leads to a 
superconducting density of states which is largely asymmetric as 
is also seen in a-b tunneling experiments of ${\sf Y\,Ba_2\,Cu_3\,O_{8-\delta}}$ 
and ${\sf Bi_2 Sr_2 Ca Cu_2 O_{8-\delta}}$. \cite{deWilde98a,renner98a} 
Fig.\ 3 (b) is a plot of the density of 
states for three doping concentrations close to the optimal value x=0.22, in the
restricted frequency range which is often used in those tunneling experiments. 

\begin{table}
\caption{Changes in the gaps for $V_0=-250$ meV, $V_1=150$ meV and $x=0.2$.}
\begin{tabular}{ddddd}
$\alpha$ & $\beta$ & $\Delta_d$ (meV) & $\Delta_s$ (meV) & $\Delta_{s*}$ (meV)\\
\tableline
0.     &   0.     &  22.1   &   0.     &  0.    \\
0.05   &   0.     &  20.5   &  -0.28   &  0.57  \\
0.     &   0.05   &  22.1   &  -0.23   &  1.54  \\
0.05   &   0.05   &  20.6   &  -0.5    &  2.01  \\
-0.05  &   0.05   &  20.4   &   0.07   &  0.86  \\
0.05   &  -0.05   &  20.4   &  -0.07   & -0.86  \\
-0.05  &  -0.05   &  20.6   &   0.5    & -2.01  
\end{tabular}
\end{table}

The main effect of an orthorhombic distortion of the square lattice is to allow 
mixing of the d and $s$-wave states, as discussed above.
Such an effect can indeed be seen in the numerical solution of the saddle point
equations, as shown in Table II. Another effects are related to the different 
role performed by the
parameters $\alpha$ and $\beta$. A finite $\alpha$ also 
reduces the value of the $d$-wave gap. On the other hand, a finite 
$\beta$ doesn't alter the $d$-wave gap so much, but induces more appreciable
$s$-wave components. The relative phase between $s$ and $d$-wave gaps on two
different twin domains can also be modeled by changing the sign of $\alpha$ and
$\beta$ as shown in the last four rows of table II. \cite{kouznetsov97a}  

The mapping of the tight-binding model onto a linearized Hamiltonian works quite
well. There are three parameters which appear naturally in the analytic 
calculation, $k_F\,\lambda$, g, and $E_d$. All of them are needed to fit
$\Delta_{s,d}^L$ onto their partners, $\Delta_{s,d}$ along the axes. The values 
extracted for 
$k_F\,\lambda/\pi$, g, and $E_d$ range from 1.21 to 1.27, 1.1 to 1.6 $eV^{-1}$,
and 205 to 480 meV, respectively.\cite{ferrer98a} 
There are no further parameters at hand so
the mapping for the remaining part of the phase diagrams can be regarded as
parameter free. Fig.\ 1 shows that the qualitative trends are indeed captured
even though the slopes of the different phase boundaries are not exactly the
same.    
\section*{Inclusion of superflow}
The Ginsburg-Landau free energy functional of a pure $d$-wave superconductor on 
a 
perfect square lattice with a superfluid velocity $\vec{v}_s=\hbar\,\vec{q}/m$, 
\cite{zapotocky97a,feder97a,joynt90a,ren95a} 
\begin{eqnarray}
f&=&\frac{h^2}{8\,\pi}\nonumber\\
&&+\frac{|\alpha|^2}{2\,\beta}\left[-\frac{1}{2}+
\frac{1}{2}(|d|-1)^2-\xi_s^2\,|(-i\,\vec{\nabla}+\vec{q}-\frac{2\,e}{\hbar\,c}
\vec{A})\,d|^2\right],\nonumber\\
\end{eqnarray}
shows that the current density 
\begin{equation}  
\vec{\j}=-2\,e\,n_s(\vec{q})\,\vec{v_s}(\vec{q})=\frac{e\,\hbar\,|\alpha|}
{m\,\L^2\,\beta}
|d(\vec{q})|^2\,\vec{q}=\frac{e\,|\alpha|}{L^2\,\beta}\,
(1-\frac{m\,v_s^2}{2\,|\alpha|})\,\vec{v}_s
\end{equation}
is completely isotropic close to the critical temperature. Here, the order
parameter $\Delta_d\propto\left[-|\alpha|/2\beta\right]^{1/2}\times 
\eta_{d,\vec{k}}
\times d$ 
and $\xi_s=\left[\hbar^2/2m|\alpha|\right]^{1/2}$ is the superconducting 
coherence length. 

\begin{figure}
\psfigurepath{plots}
\psfig{figure=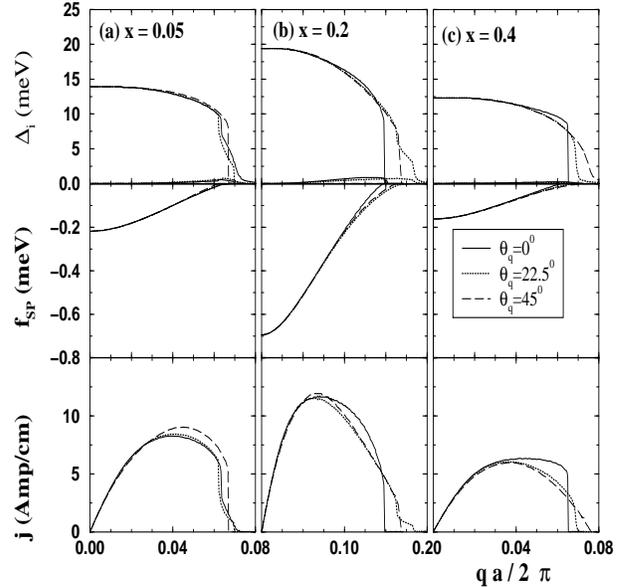,height=8cm,width=8cm,angle=270}
\vspace{0.5cm}
\caption{Zero Temperature gaps $\Delta_{s,s*,d}$, free energy per site 
$f_{SP}$, and
current density ${\rm j}$ as a function of the wave vector $q$
normalized to the size of the Brillouin zone for three different angles and
doping concentrations (a) x=0.05; (b) x=0.2; (c) x=0.4.}
\label{fig4}
\end{figure}

The relation between ${\rm j}$ and $q$ is linear at the beginning, but 
for large enough $q$, ${\rm j}$ begins to saturate and 
attains a maximum which defines the critical current ${\rm j}_c$. Afterwards, 
the  
current density decreases and eventually vanishes, although this part of the
curve corresponds to metastable states and has therefore no physical relevance. 
The linear behavior of ${\rm j}$ for small Doppler shifts 
arises from the dependence of the superfluid velocity with $q$, because $n_s$ 
is then almost constant.   
For Doppler shifts of the order of the inverse coherence length, on the
contrary, $n_s$ is largely depressed and so the current density saturates at a
maximum which defines the critical current density ${\rm j}_c$.  

This qualitative behavior displayed by superconductors close to $T_c$ is also 
manifested at low temperatures. The top panels of Fig.\ 4 show the zero 
temperature gaps as a function of $q a/2\pi$ for three different angles 
$\theta_q$ formed between $\vec{q}$ and the X axis, $0^0$, $22.5^0$ and $45^0$. 
The most distinct feature is that the $d$-wave gap follows very similar 
trends for all the angles, showing that the superflow state is indeed rather
isotropic. Besides, the gaps (a) decrease slowly as they would in the 
Ginsburg-Landau 
regime, but then drop abruptly at a large $q$ value and, finally, there is 
sometimes a strange tail; (b) small $s$ and $s^*$ components are induced and 
grow 
slowly with $q$; (c) the range of $q$-values over which the gaps are finite is 
clearly larger for the nearly optimal doping concentration x=0.2; (d) the 
$d$-wave gap is somewhat smaller for $\theta_q=0^0$ than for 
$\theta_q=22,5^0, 45^0$, but extends to larger values of $q$ for x=0.05; on the 
contrary, it is bigger for the other two doping concentrations, but drops to 
zero earlier. 

\begin{figure}
\psfigurepath{plots}
\psfig{figure=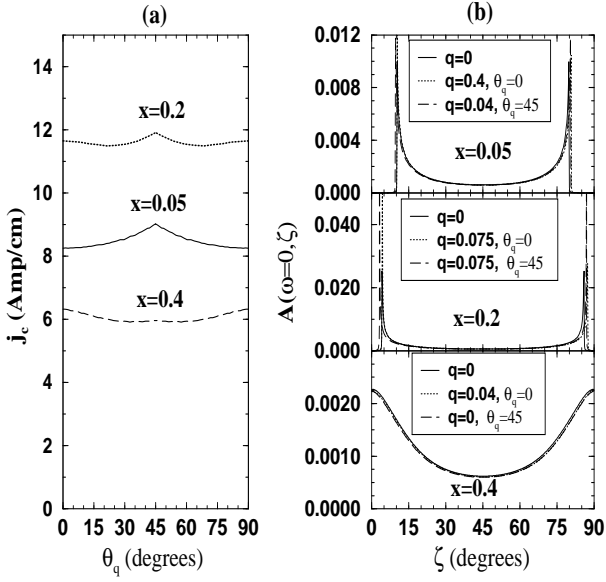,height=8cm,width=8cm}
\vspace{0.5cm}
\caption{(a) Critical current density ${\rm j}_c$ as a function of the angle 
$\theta_q$ for three doping concentrations; (b) Angular spectral function
$A(\omega=0,\zeta)$ for various values of $q$ and $\theta_q$. Top panel, x=0.05, 
middle panel, x=0.2; bottom panel, x=0.4.}
\label{fig5}
\end{figure}

The free energy is shown in the center panels. It is much the same for all
angles at each doping concentration and does not show any strange feature.
The bottom panels show the current densities. They display the Ginsburg-Landau
behavior in a loose sense, but there are again tails at the far right
range of $q$. The overall form of
the curves is fairly similar again for all the angles displayed. Moreover,
the superfluid density, which is the initial slope of the curve, seems to be
the same for all of them. The form of the curves is markedly different for the 
different doping concentrations, and the current densities attained are larger 
again for x=0.2. 

\begin{figure}
\psfigurepath{plots}
\psfig{figure=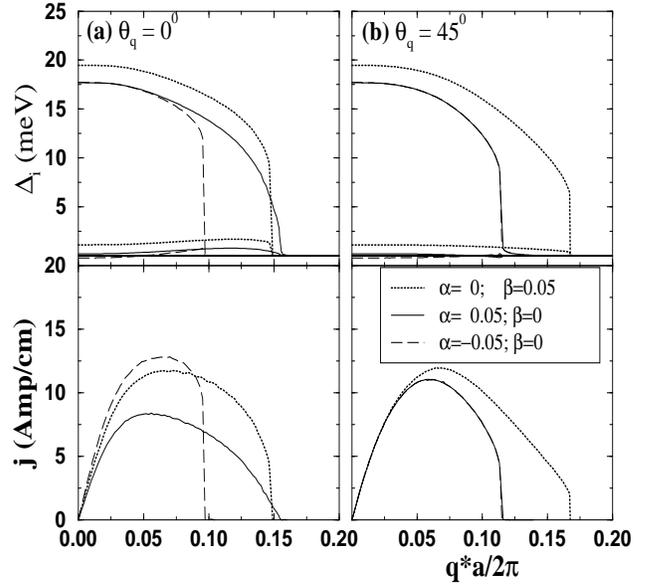,height=8cm,width=8cm}
\vspace{0.5cm}
\caption{Zero Temperature gaps $\Delta_{s,s*,d}$, and current density 
${\rm j}$ as a function of the wave vector $q$ normalized to 
the size of the Brillouin zone, for three possible values of the orthorhombic
parameters $\alpha$ and $\beta$ and a doping concentration of x=0.2. The angles
are (a) $\theta_q=0^0$ and (b) $\theta_q=45^0$.}
\label{fig6}
\end{figure}

The critical currents depend only slightly on the angle $\theta_q$, as
shown in Fig.\ 5 (a) for several doping concentrations. A closer look
reveals that ${\rm j}_c$ is maximum at $\theta_q=45^0$ and minimum at  
$\theta_q=0^0$ for x=0.05; it is still maximum at 45 degrees at x=0.2, but in
this case ${\rm j}_c$ is a non-monotonic function of the angle; the maximum
is attained at $\theta_q=0^0$ for x=0.4. This behavior can be understood in
terms of the angular spectral function
\begin{equation}
A(\omega,\zeta<45^0)=\int_0^{\pi/\cos\zeta}\,\frac{dk}{2\,\pi}\,k\,
\delta(\omega-\xi(k,\zeta)),
\end{equation}
which weights the back-flow term due to quasiparticles in Eq.\ (\ref{current}). 
$A(\omega=0,\zeta)$ (Fig.\ 5 (b)) has strong peaks for small angles when x=0.05 
and 0.2 and therefore the back-flow contribution to the current must then be 
important. $A(\omega=0,\zeta)$ is smoother for x=0.4, and accordingly, 
${\rm j}_c$
is flatter. Indeed, the angular dependence of the critical current ${\rm j}_c$
is even weaker for the linearized model; ${\rm j}_c$ is always minimum at
$\theta_q=45^0$ in such a case. Taken together, all these facts clearly
indicate that the anisotropy of the band is a more efficient source of 
anisotropy than the internal structure of the order parameter. 

The critical currents achieved are of the order of 10 Amp/cm. This leads to an
estimated critical current for ${\sf Bi_2 Sr_2 Ca Cu_2 O_{8-\delta}}$ of
$5\times 10^7\,{\rm Amp/cm^2}$, which is one order of magnitude larger than the 
value $10^6\,{\rm Amp/cm^2}$ found in thin films some time
ago.\cite{lemberger_ginsbergIII_book} Similarly, we obtain 
${\rm j}_c\simeq 2\times
10^8\,{\rm Amp/cm^2}$ for ${\sf Y Ba_2 Cu_3 O_{8-\delta}}$, while the
experimental values are of the order of $5\times 10^7\,{\rm Amp/cm^2}$, not far
away.\cite{lemberger_ginsbergIII_book}
 
The superfluid density tensor 
\begin{equation}
n_{s,i,j}=m\,\int_{\pi}^{\pi}\,\frac{d\vec{k}}{(2\,\pi)^2}\,
|v_{\vec{k}}|^2\tanh \left(\frac{\beta\,E_{\vec{k}}}{2}\right)\,
\left[M^{-1}\right]_{ij}
\end{equation}
is not proportional to $|\Delta|^2$ as happens in Ginsburg-Landau theory.

Orthorhombic effects modeled by the $\beta$ parameter do not affect the 
behavior of the $d$-wave gap or the current densities. They only induce a rather
large $s$ component, which also increases with $q$. The curves for $\Delta_d$
and ${\rm j}$ are indeed indistinguishable from those at $\beta=0$.
The impact of the other orthorhombic parameter $\alpha$ is bigger. 
Fig.\ 6 shows the gaps and current densities for two opposite
orthorhombic elongations of the X and Y axes, $\alpha=\pm0.05$. For an angle
$\theta_q=0^0$, (a) the $d$-wave gaps are initially reduced by the 
same magnitude as they should, but as $q$ increases the solution for negative 
$\alpha$ drops to zero faster; (b) the $s$-wave gaps have initially 
the same magnitude and opposite signs, but that corresponding to $\alpha=-0.05$
goes through zero, changes its sign and eventually drops to zero again; (c) the
current densities show the opposite trend, in the sense that both the 
superfluid density $n_s$ and the critical current ${\rm j}_c$ are larger for 
negative $\alpha$. Moreover, superfluid densities and critical currents for 
$\alpha=0.05$ are even smaller than those corresponding to no orthorhombic 
distortion. 

Fig.\ 6 (b) corresponds to $\theta_q=45^0$. In this case, the two orthorhombic
distortions corresponding to $\alpha=\pm 0.05$ are equivalent and the curves for 
the $d$-wave gaps fall one on top of the other. The two $s$-wave gaps have 
exactly 
the same magnitude and different sign. The current densities are also equal as 
they should. The overall trend is that a finite $\alpha$ reduces both the 
$d$-wave
gap and the critical current, although the superfluid density does not change
much.

Comparison of Figs.\ 6 (a) and (b) shows that orthorhombicity is also a more 
efficient
source of anisotropy that the internal symmetry of the order parameter function.
Indeed, not only ${\rm j}_c$ but also $n_s$ strongly depend on $\theta_q$ even 
for these small values of the parameter $\alpha$.     
       
\begin{figure}
\psfigurepath{plots}
\psfig{figure=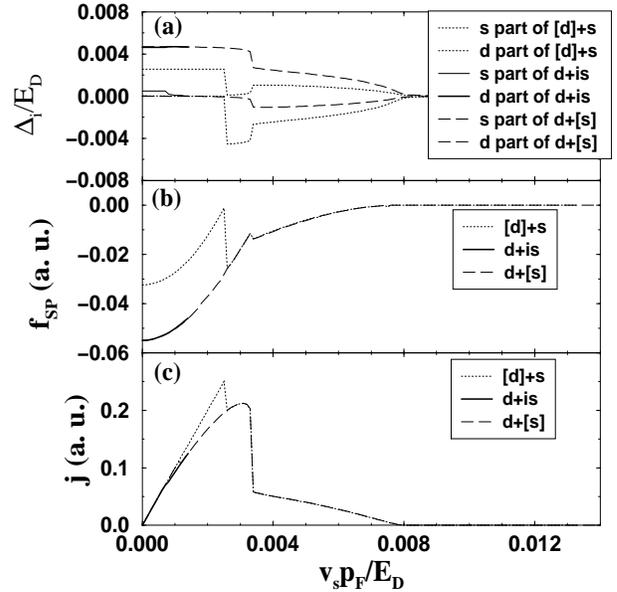,height=8cm,width=8cm}
\vspace{0.5cm}
\caption{(a) Zero temperature gaps $\Delta_i$, (b) free energy $f_{SP}$, and 
(c) current density ${\rm j}$ 
in arbitrary units as a function of the normalized superfluid velocity in the 
linearized model for $\theta_q=0^0$ and $V_s^L\,{\rm g}^L=0.15,\,V_d^L\,
{\rm g}^L=0.32$;
$d+[s]$ and $[d]+s$ indicate mixed states with dominant $d$ or $s$ components,
respectively.}
\label{fig7}
\end{figure}

The time-reversal-symmetry broken solution $d+i(s+s^*)$ to the saddle point 
equations appears for attractive $V_0$ around the phase boundary between the $d$ 
and $s+s^*$ solutions. In this part of the phase diagram, there exist not only
the $d+i(s+s^*)$ state, but also finite $s+s^*$ and d solutions. The phase
transitions from one stable state to another are therefore first order. A finite
Doppler shift gives rise to (a) mixing between the d and $s+s^*$ states so that 
each of these two solutions is of the form $s+s^*+[d]$ or $[s+s^*]+d$; where the  
subdominant component is enclosed in brackets; (b) competition between these 
two solutions and with the broken time-reversal-symmetry state.

Numerical calculations for the linearized case and suitably chosen values of 
$V_s^L$ and $V_d^L$ are displayed in Figs.\ 7 and 8 for angles $\theta_q=0^0$ and
$45^0$. They show that the  
$d+is$ state is initially more stable but the s component decreases when $v_s$ 
increases and eventually vanishes. The $d+is$ state disappears at that moment 
in favor of the $d+[s]$ state, $[d]+s$ being higher in energy. The $d+[s]$ 
solution becomes metastable once the critical current density of that state
is reached. There is then a phase transition to the $[d]+s$ state, which has a
higher ${\rm j}_c$, because it is the current what is actually imposed in an 
experimental setup. Above this second critical current density, the system
reverts to its normal phase. 
 
\begin{figure}
\psfigurepath{plots}
\psfig{figure=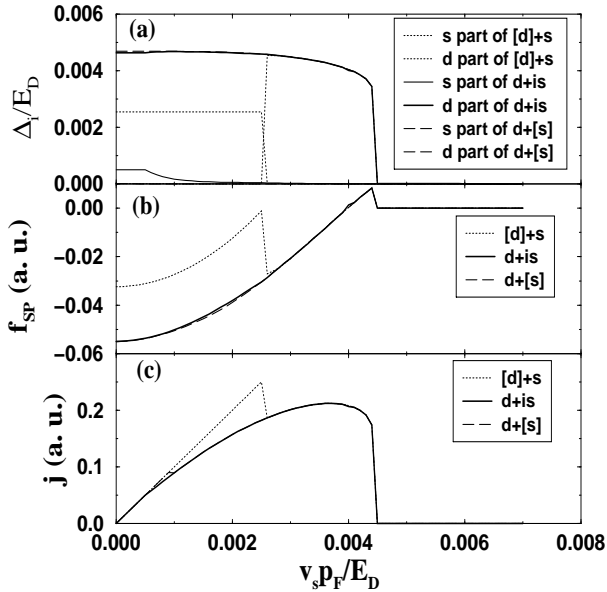,height=8cm,width=8cm}
\vspace{0.5cm}
\caption{(a) Zero temperature gaps $\Delta_i$, (b) free energy $f_{SP}$ ,and 
(c) current density ${\rm j}$
in arbitrary units as a function of the normalized superfluid velocity in the 
linearized model for $\theta_q=45^0$ and $V_s^L\,{\rm g}^L=0.15,\,
V_d^L\,{\rm g}^L=0.32$.}
\label{fig8}
\end{figure}

\section*{Conclusions}
The formalism needed to include superflow in a gauge invariant way in a 
superconductor has been discussed in detail and shown to be equivalent to give 
a Doppler shift to the wave vectors in the kinetic energy part of the 
Hamiltonian. This has been applied to study supercurrents in a clean, infinite, 
$d$-wave superconductor on an orthorhombically distorted square lattice, using a
model which describes some of the features of the superconducting state of
several High-temperature superconductors. The behavior of gaps, free energy, 
current density and critical current as a function of modulus and angle of the
Doppler shift, doping, and orthorhombic distortion have been discussed.

There are three possible sources of anisotropy in the discussed model: the 
internal symmetry of the
order parameter, the complex dispersion relation of the band and the
orthorhombic distortion of the lattice. The results of this article show that
the most important one is orthorhombicity distortion, then the form of the band
and last and least the $d$-wave form of the gap function.

After this work was completed we found that D. Feder and C. Kallin
had also been studying the behavior of supercurrents in $d$-wave superconductors
and had reached conclusions which agree with those of this article.   

\acknowledgments

It is a pleasure to thank F. Sols, A. D. Zaikin and D. Feder for useful 
conversations, as well as financial support from the Spanish Direcci\'on 
General de Ense\~nanza Superior, Project No. PB96-0080-C02 and 
the TMR Program of the European Union, contract No. FMRX-CT96-0042.

\bibliographystyle{prsty}

\begin{thebibliography}{10}

\bibitem{reitz_book}
J.~R. Reitz, F.~J. Milford, and R.~W. Christy, {\em Foundations of
  Electromagnetic Theory} (Addison-Wesley, Reading, Massachussets, 1979).

\bibitem{pyun89a}
D.~S. Pyun, E.~R. Ulm, and T.~R. Lemberger, Phys. Rev. B {\bf 39},  4140
  (1989).

\bibitem{schrieffer_book}
J.~R. Schrieffer, {\em Theory of superconductivity} (Addison-Wesley, Reading,
  Massachussets, 1964).

\bibitem{ferrer_thesis}
J. Ferrer, Ph.D. thesis, Universidad Aut\'onoma de Madrid, 1990.

\bibitem{furusaki91a}
A. Furusaki and M. Tsukada, Solid State Commun. {\bf 78},  299  (1991).

\bibitem{kim_parks_book}
Y.~B. Kim and M.~J. Stephen,  in {\em Superconductivity, vol. II}, edited by
  R.~D. Parks (Dekker, New York, 1969).

\bibitem{dessau93a}
D.~S. Dessau {\it et~al.}, Phys. Rev. Lett. {\bf 71},  2781  (1993).

\bibitem{gofron94a}
K. Gofron {\it et~al.}, Phys. Rev. Lett. {\bf 73},  3302  (1994).

\bibitem{norman95a}
M.~R. Norman, M. Randeria, H. Ding, and J.~C. Campuzano, Phys. Rev. B {\bf 52},
   615  (1995).

\bibitem{ferrer98a}
J. Ferrer, M.~A. Gonz\'alez-Alvarez, and J.~S\'anchez-Ca\~nizares, Phys. Rev.
  B {\bf 57},  7470  (1998).

\bibitem{zapotocky97a}
M. Zapotocky, D. Maslov, and P.~M. Goldbart, Phys. Rev. B {\bf 55},  6599
  (1997).

\bibitem{maki_parks_book}
K. Maki,  in {\em Superconductivity, vol. II}, edited by R.~D. Parks (Dekker,
  New York, 1969).

\bibitem{bagwell94a}
P. Bagwell, Phys. Rev. B {\bf 49},  6841  (1994).

\bibitem{canizares97a}
J.~S\'anchez-Ca\~nizares and F. Sols, Phys. Rev. B {\bf 55},  531  (1997).

\bibitem{martin95a}
A. Martin and C.~J. Lambert, Phys. Rev. B {\bf 51},  17999  (1995).

\bibitem{otterlo97a}
A. van Otterlo, D.~S. Golubev, A.~D. Zaikin, and G. Blatter, condmat/9703124
  (unpublished)  .

\bibitem{ding97a}
H. Ding {\it et~al.}, condmat/9712100 (unpublished)  .

\bibitem{scalapino95a}
D.~J. Scalapino, Physics Reports {\bf 250},  330  (1995).

\bibitem{chubukov96a}
A.~V. Chubukov, D. Pines, and B.~P. Stojkovic, J. Phys. Condens. Matter {\bf
  8},  10017  (96).

\bibitem{annett_ginsbergV_book}
J. Annett, N. Goldenfeld, and A.~J. Leggett,  in {\em Physical Properties of
  High Temperature superconductors V}, edited by D.~M. Ginsberg (World
  Scientific, New Jersey, 1996).

\bibitem{micnas90a}
R. Micnas, J. Ranninger, and S. Robaszkiewicz, Rev. Mod. Phys. {\bf 62},  113
  (1990).

\bibitem{nazarenko96a}
A. Nazarenko, A. Moreo, E. Dagotto, and J. Riera, Phys. Rev. B {\bf 54},  R768
  (1996).

\bibitem{feder97a}
D.~L. Feder and C. Kallin, Phys. Rev. B {\bf 55},  559  (1997).

\bibitem{bealmonod96a}
M.~T. Beal-Monod and K. Maki, Phys. Rev. B {\bf 53},  5775  (1996).

\bibitem{musaelian95a}
K.~A. Musaelian, J. Betouras, A.~V. Chubukov, and R. Joynt, condmat/9507085
  (unpublished)  .

\bibitem{odonovan95a}
C. O'Donovan and J.~P. Carbotte, condmat/9502035 (unpublished)  .

\bibitem{chandra88a}
P. Chandra and B. Dou\c{c}ot, Phys. Rev. B {\bf 38},  9335  (1988).

\bibitem{ferrer93a}
J. Ferrer, Phys. Rev. B {\bf 47},  8769  (1993).

\bibitem{duffy97a}
D. Duffy {\it et~al.}, Phys. Rev. B {\bf 56},  5597  (1997).

\bibitem{spathis92a}
P.~N. Spathis, M.~P. Soerensen, and N. Lazarides, Phys. Rev. B {\bf 45},  7360
  (1992).

\bibitem{miyakawa98a}
N. Miyakawa {\it et~al.}, Phys. Rev. Lett. {\bf 80},  157  (1998).

\bibitem{deWilde98a}
Y. DeWilde {\it et~al.}, Phys. Rev. Lett. {\bf 80},  153  (1998).

\bibitem{renner98a}
Ch. Renner {\it et~al.}, Phys. Rev. Lett. {\bf 80},  149  (1998).

\bibitem{kouznetsov97a}
K.~A. Kouznetsov {\it et~al.}, Phys. Rev. Lett. {\bf 79},  3050  (1997).

\bibitem{joynt90a}
R. Joynt, Phys. Rev. B {\bf 41},  4271  (1990).

\bibitem{ren95a}
Y. Ren, J.H. Xu, and C.~S. Ting, Phys. Rev. Lett. {\bf 74},  3680  (1995).

\bibitem{lemberger_ginsbergIII_book}
T.~R. Lemberger,  in {\em Physical Properties of High Temperature
  superconductors III}, edited by D.~M. Ginsberg (World Scientific, New Jersey,
  1992).

\end{thebibliography}

\end{document}